\documentclass[prd,aps,preprint,nofootinbib,superscriptaddress]{revtex4}
\usepackage{epsfig}
\usepackage{amsmath}
\input{epsf}

\begin{document}


\vspace*{2cm}
\title{Next-to-leading Order Calculation of the
Single Transverse Spin Asymmetry in the Drell-Yan Process \\[1cm]}

\author{Werner Vogelsang}
\email{vogelsan@quark.phy.bnl.gov}
\affiliation{Physics Department, Brookhaven National Laboratory,
             Upton, NY 11973}
\author{Feng Yuan}
\email{fyuan@quark.phy.bnl.gov}
\affiliation{Nuclear Science Division,
Lawrence Berkeley National Laboratory, Berkeley, CA
94720}\affiliation{RIKEN BNL Research Center, Building 510A,
             Brookhaven National Laboratory, Upton, NY 11973}
\date{\today\\[1cm]}

\begin{abstract}
We calculate the next-to-leading order perturbative QCD
corrections to the transverse momentum weighted
single transverse spin asymmetry in Drell-Yan lepton pair production
in hadronic collisions. We identify the splitting function relevant for
the scale evolution of the twist-three quark-gluon correlation function.
We comment on the consequences of our results for phenomenology.
\end{abstract}

\maketitle

\newcommand{\nn}{\nonumber}
\newcommand{\be}{\begin{equation}}
\newcommand{\ee}{\end{equation}}
\newcommand{\ben}{\[}
\newcommand{\een}{\]}
\newcommand{\beqn}{\begin{eqnarray}}
\newcommand{\eeqn}{\end{eqnarray}}
\def\beq{\begin{equation}}
\def\eeq{\end{equation}}
\def\beeq{\begin{eqnarray}}
\def\eeeq{\end{eqnarray}}
\newcommand{\ba}{\begin{array}}
\newcommand{\ea}{\end{array}}
\newcommand{\Tr}{{\rm Tr} }
\newcommand{\dsp}{\displaystyle}
\newcommand\Eqn[1]{Eq.~(\ref{#1})}  
\newcommand\eqn[1]{(\ref{#1})}      

\section{Introduction}
Single transverse spin asymmetries (SSAs) in high energy hadronic
reactions continue to attract much theoretical and experimental
interest. They are defined as differences of cross sections
when one of the initial hadrons' transverse spin is flipped, divided
by the sum: $A_N\sim
(d\sigma(S_\perp)-d\sigma(-S_\perp))/(d\sigma(S_\perp)+d\sigma(-S_\perp))$.
The theoretical description of SSAs has proven to be a
challenge~\cite{review}, since the leading collinear partonic contribution
to the asymmetries vanishes~\cite{KPR}. Over the past few years, there have
been a number of theoretical developments that have led to much progress
in the exploration of the underlying physics for single spin asymmetry
phenomena. These developments mainly follow two lines: the so-called
transverse momentum dependent (TMD)
approach~\cite{Brodsky:2002cx,Collins:2002kn,Ji:2002aa,Boer:2003cm,
JiMaYu04,ColMet04,Sivers:1990fh,Anselmino:1994tv}, which uses parton
distributions and/or fragmentation functions that depend on partonic
transverse momentum, and the twist-three quark-gluon correlation function
approach~\cite{Efremov,qiu,new,Kanazawa:2000hz,Boer:2001tx,{new1},Eguchi:2006mc,Koike:2009ge}.
More recently, it has been found in some cases that the two approaches
are closely related and describe the same physics~\cite{jqvy,bbdm}.

So far, however, phenomenological applications of the approaches
have been limited to the ``bare'' parton model, that is, to the zeroth
order of perturbation theory without any QCD corrections, as the latter
were generally
not available. This situation was remedied very recently when the
leading-order (LO) kernels for the scale evolution of the relevant
twist-three correlation functions were derived~\cite{kang,zhou}. In this paper,
we take a further step toward a more comprehensive QCD description of
single-spin phenomena by calculating next-to-leading order (NLO)
QCD corrections for a particular observable, the transverse momentum weighted
SSA in Drell-Yan lepton pair production.

As demonstrated by many examples, next-to-leading order perturbative QCD
(pQCD) corrections are typically very important in hadronic processes. They
often lead to significant $K$-factors, and also allow estimates
of the size of yet higher order corrections. Moreover, an NLO calculation
for a particular physical process will provide a direct test of QCD
factorization for the associated observable, complementing the general
arguments for such factorizations~\cite{JiMaYu04,ColMet04,qiu-fac}.
One-loop pQCD corrections to the hard-scattering
factors in the TMD factorization
approach have been calculated for the observables associated with the
so-called $k_\perp$-even TMD parton distributions~\cite{JiMaYu04}.
For the related $k_\perp$-odd TMD observables, especially relevant for
SSAs in various processes, there has not been any particular calculation
so far. The same is true for the twist-three approach, where all
calculations for the SSAs so far have been at leading order
only~\cite{qiu,new,new1}. Previous studies have shown that QCD corrections
for higher-twist observables are much more complicated to obtain
than for leading-twist ones~\cite{Ratcliffe:1985mp,Ali:1991em,Belitsky:1997zw,{belitsky}}.
This is related both to technical difficulties resulting from more complex partonic
states, and to possible mixing between higher-twist matrix
elements~\cite{Ratcliffe:1985mp}.

On the other hand,
the recent developments, especially the consistency between the TMD
approach and the twist-three approach found in~\cite{jqvy}, have provided
confidence in our understanding of the underlying theoretical description
of single-spin phenomena. They naturally motivate a study of NLO corrections
to SSA observables.
The calculations and results of~\cite{jqvy} will be the starting point
for our derivation of the NLO corrections to the SSA in the Drell-Yan
process. In this process, a transversely polarized nucleon
with momentum $P_A$ scatters off an unpolarized nucleon ($P_B$) to produce
a virtual photon with invariant mass $Q$ and transverse momentum $q_\perp$,
which subsequently decays into a lepton pair,
\begin{equation}
p_\uparrow (P_A,S_\perp) \, p(P_B)\to \gamma^*(Q^2,q_\perp)+X
\to\ell^+\ell^-+X \ ,
\end{equation}
where $S_\perp$ is the transverse polarization vector of the incident nucleon.
The Drell-Yan process is bested suited as a first example for the
calculation of NLO corrections to single-spin processes. It is
among the simplest processes in hadronic scattering, and its single
spin asymmetry is also kinematically simpler than those for other processes.
For example, for scattering with single transverse polarization,
the pair transverse momentum $q_\perp$ and the polarization vector
$S_\perp$ are simply correlated as $\epsilon^{\alpha\beta}S_\perp^\alpha
q_\perp^\beta=|S_\perp||q_\perp|\sin\phi$, where $\phi$ is the azimuthal
angle of $\vec{q}_\perp$ relative to that of $\vec{S}_\perp$. If
$\vec{q}_\perp$ is measured experimentally, the corresponding
single spin asymmetry receives contributions from the so-called Sivers
effect in the TMD approach, applicable when $q_\perp\ll Q$,
or from the twist-three Qiu-Sterman matrix elements when $q_\perp\sim Q$.
As we mentioned above, the two approaches coincide in the kinematic
regime of overlap~\cite{jqvy}. In the following, we will make use of
this fact.

The transverse momentum of the virtual photon (or the lepton pair)
generally depends on various transverse momenta in the process, namely
those of the initial partons, and those generated by gluon radiation.
For the cross section differential in transverse momentum, one has to
be careful to classify the different contributions, as a TMD factorization
only exists in the limit of small transverse momentum,
$q_\perp\ll Q$~\cite{JiMaYu04,{jqvy}}. However, if we integrate over
all transverse momentum $q_\perp$, the cross section will depend only on
the {\it longitudinal} momentum fraction of the virtual photon, and a
collinear factorization approach will apply. For the single transverse-spin
dependent cross section, we have to suitably weigh with transverse momentum
in order to obtain a non-vanishing result, because the unintegrated cross
section has linear dependence on $\vec{q}_\perp$. The weighted cross section
is defined as~\cite{boer98,{boer-dy}}:
\begin{equation}
\langle q_\perp\Delta \sigma (S_\perp) \rangle\equiv \int d^2q_\perp
|q_\perp|\sin\phi \frac
{d\Delta \sigma(S_\perp)}{d^2q_\perp} \ ,
\end{equation}
where we have simplified the expression by omitting dependence on any
other kinematic variables. Since the transverse momentum has been integrated
out, the above weighted cross section can be properly formulated in the
collinear factorization approach~\cite{bbdm}, 
and can be factorized into parton
distributions and/or twist-three correlation functions for the incident
nucleons, and partonic hard-scattering functions. In case of the SSA,
the quark-gluon correlation function in the polarized nucleon will be an
important ingredient in the factorization formula. It will be part of the
following factorization formula for the above $q_\perp$-weighted cross
section:
\begin{equation}
\frac{d\langle q_\perp\Delta \sigma (S_\perp) \rangle}{dQ^2}=
\sigma_0\int\frac{dx_1}{x_1}\frac{dx_2}{x_2}
\frac{dx'}{x'} T_{F,q}(x_1,x_2)\bar q(x')
{\cal H}(x_1,x_2;x') \ ,
\end{equation}
where $\sigma_0 = 4\pi\alpha_{\rm em}^2/3N_C sQ^2$, with
$s=(P_A+P_B)^2$,
$\bar q(x')$ denotes the anti-quark distribution of the unpolarized nucleon,
and $T_{F,q}$ the Qiu-Sterman matrix element for quark $q$.
We have restricted ourselves here to one quark flavor; extension to
more flavors and to scattering off a quark from the unpolarized nucleon
is trivial. In the following, we will drop the label $q$ of $T_{F,q}$
for simplicity.
As indicated, $T_F$ is a function of two separate light-cone
variables, and thus the convolution over momentum fraction will
include both, as we will see. $T_F$ is defined as
\begin{eqnarray}
T_F(x_1,x_2) &\equiv & \int\frac{d\zeta^-d\eta^-}{4\pi} e^{i(x_1
P_A^+\eta^-+(x_2-x_1)P_A^+\zeta^-)}
\nonumber \\
&\times & \epsilon_\perp^{\beta\alpha}S_{\perp\beta} \,
\left\langle P_A,S|\overline\psi(0){\cal L}(0,\zeta^-)\gamma^+
g{F_\alpha}^+ (\zeta^-) {\cal L}(\zeta^-,\eta^-)
\psi(\eta^-)|P_A,S\right\rangle  \ ,  \label{TF}
\end{eqnarray}
where ${\cal L}$ is the proper gauge link to make the matrix
element gauge invariant, and where the sums over color and spin
indices are implicit.

In the factorization formula Eq.~(3), the hard-scattering function can
be expanded as a series in the strong coupling constant,
\begin{equation}
{\cal H}={\cal H}^{(0)}+\frac{\alpha_s}{2\pi}{\cal H}^{(1)}+\cdots \ ,
\end{equation}
where ${\cal H}^{(0)}$ is the leading order term, ${\cal H}^{(1)}$
the NLO one, and so forth. In the
following, we will demonstrate that the above factorization formula
is valid at NLO level. In particular, the collinear divergence
can be factorized into the parton distribution and quark-gluon correlation
function, whereas the hard coefficient function is free of any divergence.
The real-gluon radiation diagrams have already been studied in~\cite{jqvy},
and the results can be carried over to our present calculation with
relatively little effort. We will compute the virtual corrections
as well. It is important to check that the soft divergence in
real-gluon radiation is canceled by that in the virtual diagrams,
so that we are left with only collinear divergences, which can be
absorbed into the parton distribution and/or the twist-three correlation
function, where they give rise to the scale evolution of the distributions.

The rest of the paper is organized as follows. In Sec. II, we will derive the
leading order expression for the hard coefficient, and calculate the
virtual correction at next-to-leading order. In Sec. III, we calculate the
real-gluon radiation contributions, and combine them with the virtual
corrections. We will show that the soft divergence is canceled in the sum,
and that the remaining collinear divergence can be removed by collinear
factorization. We conclude our paper in Sec. IV.

\section{Born Diagrams and Virtual Corrections}

At the leading order, the virtual photon is produced in the
quark-antiquark annihilation subprocess. In order to obtain a non-vanishing
weighted transverse-spin-dependent cross section
$\langle q_\perp \Delta \sigma(S_\perp)\rangle$, we have to include
an initial state interaction as shown in Fig.~1, which provides the
required phase~\cite{Boer:2001tx}. We perform our calculations
in covariant gauge.  Let $p'$ denote the momentum of the
incident anti-quark, $k_{q1}$ that of the initial quark to the left
of the cut, $k_{q2}$ that on the right, and $k_g=k_{q2}-k_{q1}$
the momentum of the polarized gluon attaching to the hard part.
This attachment may take place on the left side of the cut, as shown
in Fig.~1(a), or on the right side, as in 1(b).

\begin{figure}[t]
\begin{center}
\includegraphics[width=12cm]{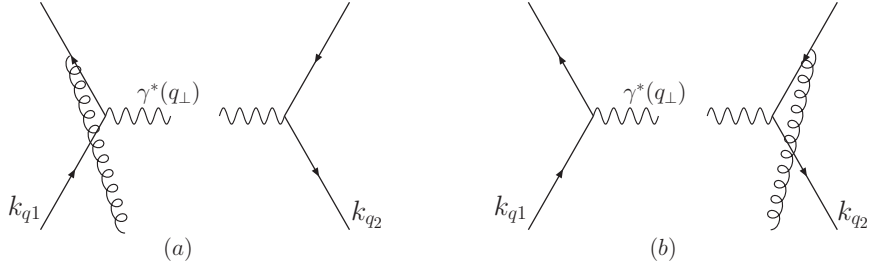}
\end{center}
\vskip -0.4cm \caption{\it Leading order contribution to the weighted
transverse spin-dependent cross section.}
\end{figure}

The polarized gluon is associated with a
gauge potential $A^\mu$, and one of the leading contributions
comes from its component $A^+$. The gluon's momentum is dominated
by $x_gP+k_{g\perp}$, where $x_g$ is the longitudinal momentum
fraction with respect to the polarized proton. The transverse
momentum $k_{g\perp}$ flows through the perturbative diagram and
returns to the polarized proton through the quark lines. The
contribution to the single-transverse-spin asymmetry arises from
terms linear in $k_{g\perp}$ which, when combined with $A^+$,
yield $\partial^\perp A^+$, a part of the gauge field strength
tensor $F^{\perp +}$. In order to compute this contribution, we
expand the partonic scattering amplitudes in terms of $k_{g\perp}$
up to the linear term. The weighted cross section can, in general,
be written as
\begin{equation}
\frac{d\langle q_\perp \Delta\sigma(S_\perp)\rangle}{dQ^2}=
\frac{\epsilon^{\beta\alpha}S_\perp^\beta}{2s}
\int\frac{d^4k_{q1}}{(2\pi)^4}\frac{d^4k_{q2}}{(2\pi)^4}
\left( q_\perp^\alpha H(k_{q1},k_{q2};Q^2)\right)
\bar q(x') T_a(k_{q1},k_{q2}) \ ,
\end{equation}
where $T_a(k_{q1},k_{q2})$ represents the non-perturbative matrix element
for the polarized nucleon with full momentum dependence on $k_{q1}$ and
$k_{q2}$.
In the above weighted cross section, we have an explicit term
$q_\perp^\alpha$ in the integral, along with the hard partonic
part $H(k_{q1},k_{q2};Q^2)$.

To obtain the collinear factorization formula Eq.~(3), we have to perform
a collinear expansion of the hard partonic part. For the leading
Born diagrams in Fig.~1, we find that $q_\perp$ is related to the
transverse momenta of the two quark lines as:
$q_\perp=k_{q2\perp}$ for Fig.~1(a) and $q_\perp=k_{q1\perp}$
for Fig.~1(b). Therefore, the contribution from Fig.~1(a) to
the collinear expansion of Eq.~(6) will be
\begin{equation}
\left(q_\perp^\alpha H(k_{q1},k_{q2};Q^2)\right)|_{\rm Fig.1(a)}=
\frac{ig}{-(k_{q2}^+-k_{q1}^+)-i\epsilon}k_{q2\perp}^\alpha\ ,
\label{eq7}
\end{equation}
where the propagator associated with the initial state interaction
produces the pole at $k_{g}^+=0$. The single spin asymmetry arises
from the phase of this pole.
Similarly, the contribution from Fig.~1(b) will be
\begin{equation}
\left(q_\perp^\alpha H(k_{q1},k_{q2};Q^2)\right)_{\rm Fig.1(b)}=
-\frac{ig}{-(k_{q2}^+-k_{q1}^+)-i\epsilon}k_{q1\perp}^\alpha\ .
\label{eq8}
\end{equation}
The total contribution is thus
\begin{equation}
\left(q_\perp^\alpha H(k_{q1},k_{q2};Q^2)\right)_{\rm Fig.1(a+b)}=
\frac{ig}{-(k_{q2}^+-k_{q1}^+)-i\epsilon}
\left(k_{q2\perp}^\alpha-k_{q1\perp}^\alpha\right)
=\frac{ig}{-(k_{q2}^+-k_{q1}^+)-i\epsilon}k_{g\perp}^\alpha\ .
\end{equation}
When integrated over the transverse and light-cone-minus components of the
two momenta $k_{q1}$ and $k_{q2}$, the combined terms $T_a(k_{q1},k_{q2})$
and  $k_{g\perp}$ produce the matrix element $T_F(x,x)$. One then
obtains the leading order contribution to the weighted cross section
as~\cite{boer98}
\begin{equation}
\frac{d\langle q_\perp\Delta \sigma (S_\perp) \rangle}{dQ^2}=
\sigma_0\int\frac{dx}{x}\frac{dx'}{x'} T_F(x,x)\bar q(x')
\delta (1-Q^2/xx's) \ ,
\end{equation}
from which one can readily determine the leading order hard coefficient to be
\begin{equation}
{\cal H}(x_1,x_2;x')=\delta(1-\frac{x_2}{x_1})\delta (1-z) \ ,
\end{equation}
where $z=Q^2/\hat s$ with $\hat s=x_1x' s$.

The above derivation shows that the Born kinematics greatly simplify
the collinear expansion for the hard partonic part, because of momentum
conservation. We can utilize this feature in the calculations of the
virtual corrections to the Born diagrams as well.
At one-loop order, the virtual corrections contain two types of diagrams shown
in Fig.~2. In the upper two diagrams $(a,b)$ of Fig.~2 the polarized gluon attaches to
the side of the cut opposite from the loop correction which is
represented by a blob. On the other hand, in the lower two diagrams $(c,d)$,
the gluon attaches to the side that also has the loop.
The loops are displayed in detail in Figs.~3 and 4.
Fig.~3 is the usual vertex correction, plus self-energy diagrams. If
the gluon is on the side of the loop, we have to attach
the gluon to all possible places in the virtual diagrams,
see Fig.~4.

\begin{figure}[t]
\begin{center}
\includegraphics[width=11cm]{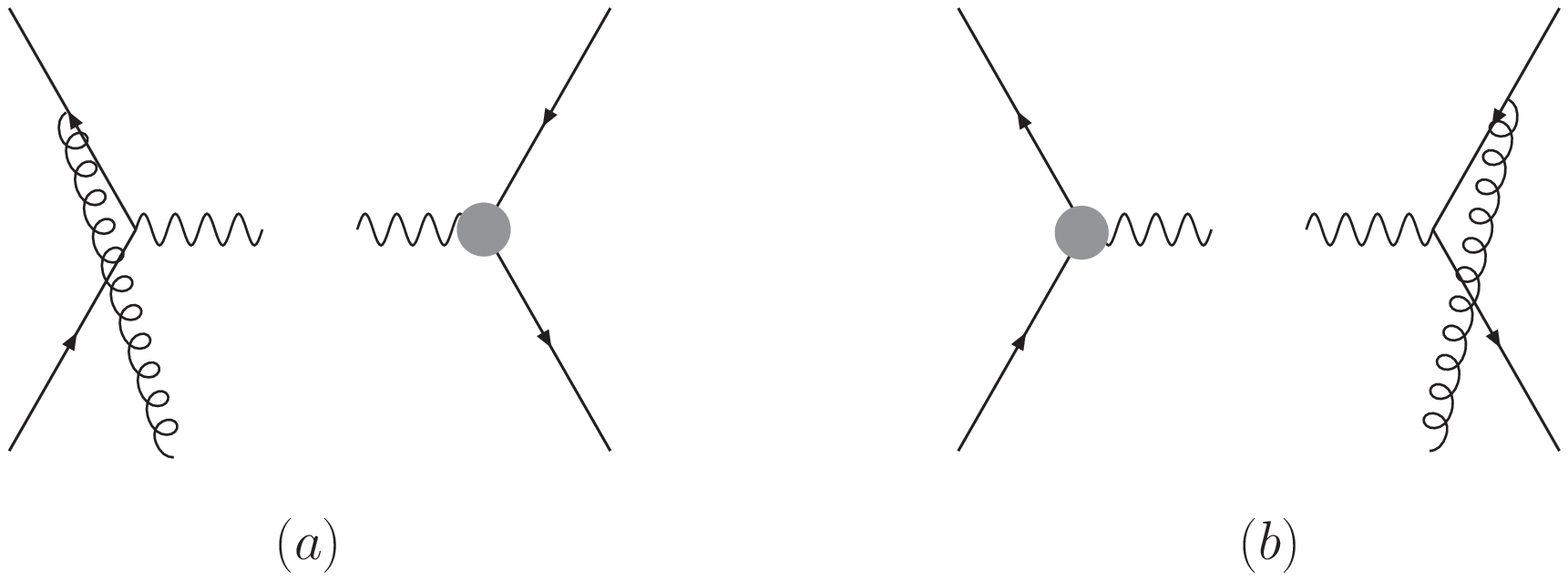}
\includegraphics[width=11cm]{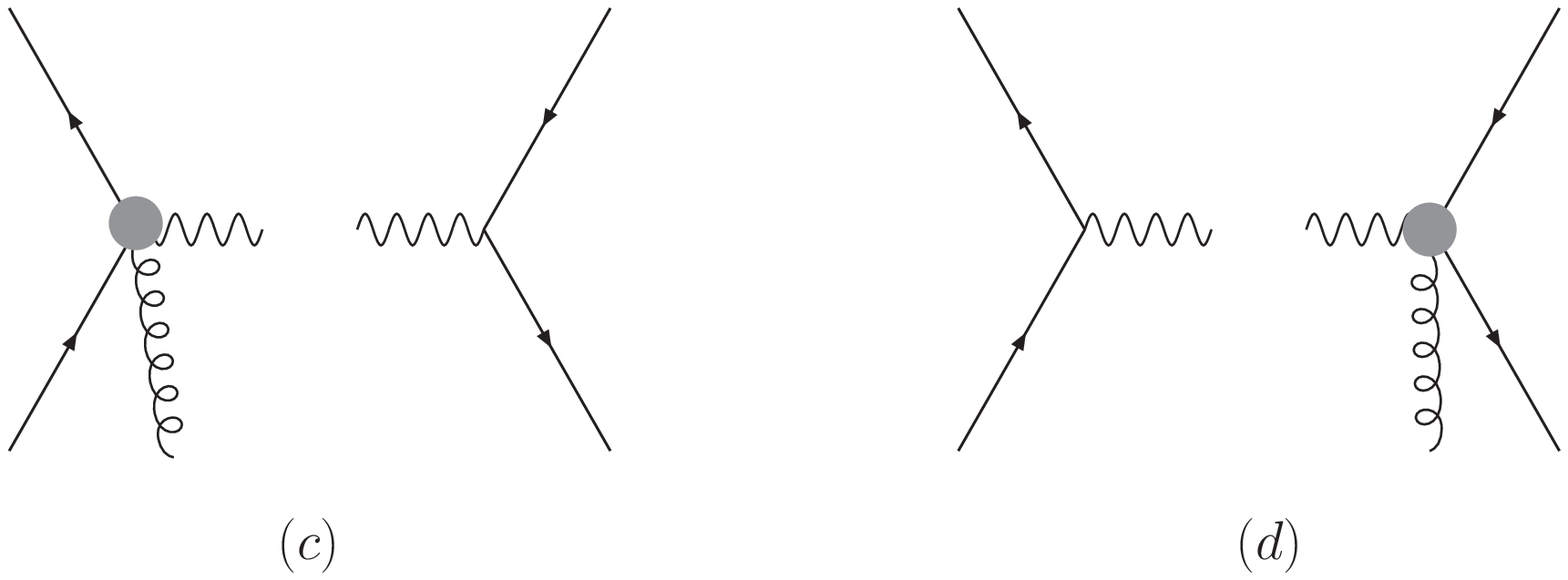}
\end{center}
\vskip -0.4cm \caption{\it One-loop virtual correction to the
weighted cross section: the gluon attaches to the opposite side of
the loop correction (upper two diagrams);
the gluon attaches to the same side of the loop
corrections (lower two diagrams).\label{f2}}
\end{figure}

\begin{figure}[t]
\begin{center}
\includegraphics[width=11cm]{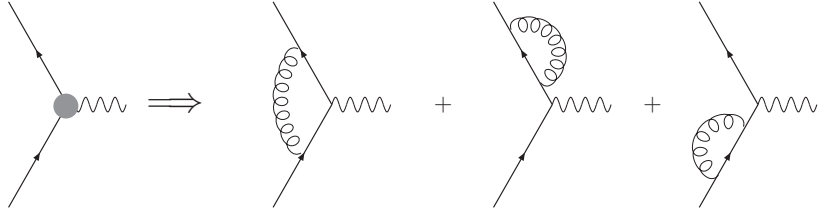}
\end{center}
\vskip -0.4cm \caption{\it Corrections to the quark-antiquark-photon
vertex, corresponding to the blob in the upper two diagrams of Fig.~\ref{f2}.}
\end{figure}

\begin{figure}[t]
\begin{center}
\includegraphics[width=13cm]{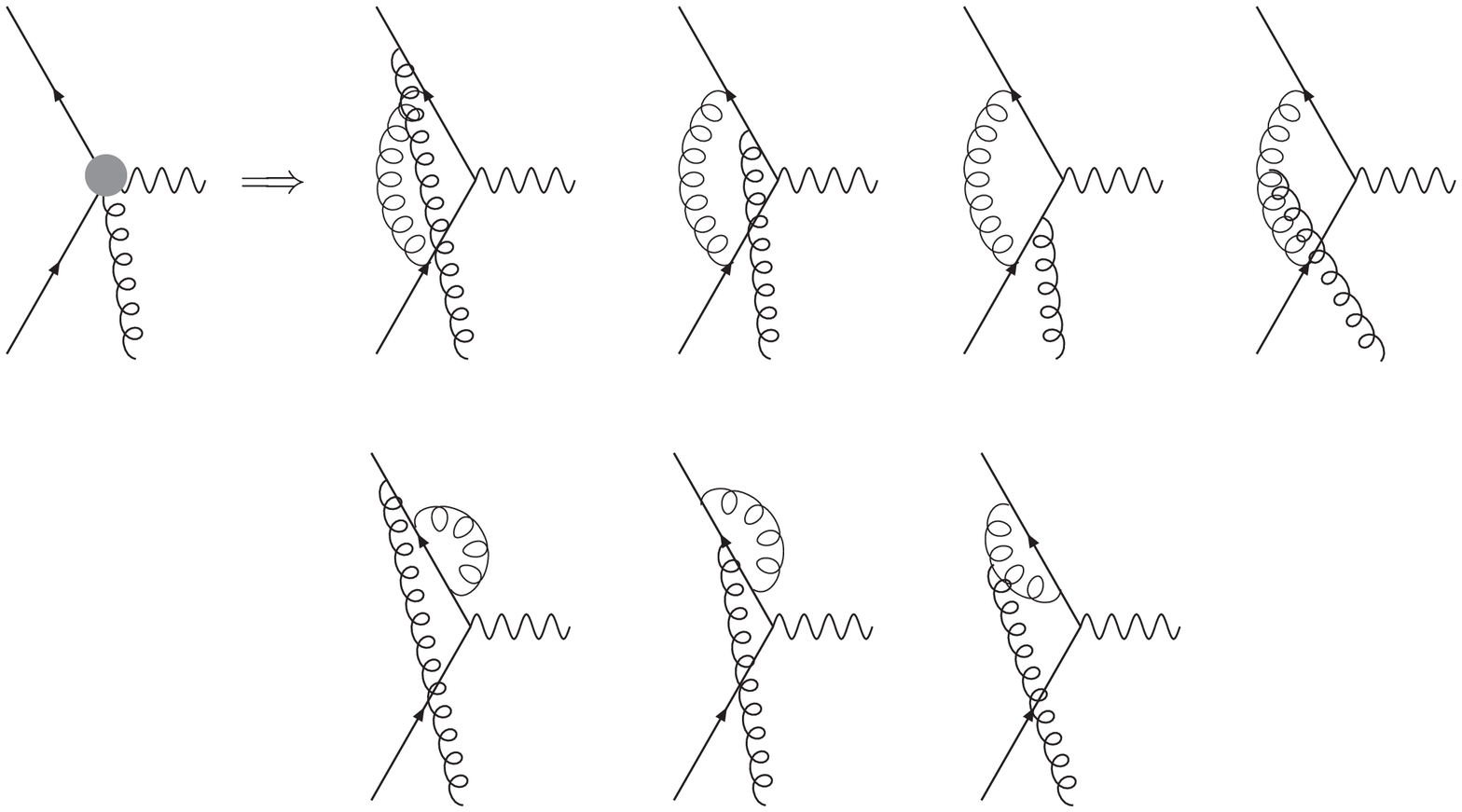}
\end{center}
\vskip -0.4cm \caption{\it Corrections to the quark-antiquark-photon
vertex with gluon attachment, corresponding to the blob in the lower
two diagrams of Fig.~\ref{f2}.}
\end{figure}

As before, the Born kinematics simplifies the collinear expansion in the
calculations of these diagrams. For example, to obtain the contributions
from Figs.~2(a,b), we can simply multiply the Born
result in Eq.~(9) by the known~\cite{dy} Drell-Yan virtual correction
factor, giving
\begin{equation}
\left(q_\perp^\alpha H(k_{q1},k_{q2};Q^2)\right)|_{\rm Fig.2(a+b)}=
\frac{ig}{-(k_{q2}^+-k_{q1}^+)-i\epsilon}k_{g\perp}^\alpha
\frac{\alpha_s}{4\pi}C_F\left(\frac{4\pi\mu^2}{Q^2}\right)^\epsilon
\left[-\frac{2}{\epsilon^2}-\frac{3}{\epsilon}-8+\pi^2\right] \ ,
\label{eq12}
\end{equation}
where $C_F=(N_c^2-1)/2N_c$ with $N_c=3$ the number of colors.
Here we have used dimensional regularization,
with $D=4-2\epsilon$ space-time dimensions and $\mu$ the mass scale
to be introduced in order to keep coupling constants
dimensionless~\footnote{Note that we also follow Ref.~\cite{dy} to
absorb a factor $(1-\epsilon)$ in the normalization $\sigma_0$,
which universally appears in all matrix elements and hence does not
affect the final results.}.
In obtaining this result,
it is essential that the one-loop virtual correction in the Drell-Yan
process amounts to a simple multiplicative factor to the vertex
$\gamma^\mu$. As for the Born diagram, the collinear expansion is trivial,
and the phase for the SSA comes from the initial state interaction,
i.e., the denominator of Eq.~(\ref{eq12}).

The calculation of Figs.~2(c,d) is more cumbersome,
but again the collinear expansion will receive contributions proportional
to $k_{q1\perp}$ and $k_{q2\perp}$ just as in Eqs.~(\ref{eq7}), (\ref{eq8}).
After a lengthy calculation, we find that the diagrams in Fig.~4
lead to the following result for the initial state interaction
contribution:
\begin{equation}
\left(q_\perp^\alpha H(k_{q1},k_{q2};Q^2)\right)|_{\rm Fig.2(c+d)}=
\frac{ig}{-(k_{q2}^+-k_{q1}^+)-i\epsilon}k_{g\perp}^\alpha
\frac{\alpha_s}{4\pi}C_F\left(\frac{4\pi\mu^2}{Q^2}\right)^\epsilon
\left[-\frac{2}{\epsilon^2}-\frac{3}{\epsilon}-8+\pi^2\right] \ ,
\label{eq13}
\end{equation}
identical to that for Figs.~2(a,b).
Substituting the results in Eqs.~(\ref{eq12}) and (\ref{eq13})
into the collinear expansion formula, and accounting for the leading-order
phase space in $D$ dimensions, we find the total virtual
correction to the weighted single-spin cross section:
\begin{eqnarray}
\sigma_0\frac{\alpha_s}{2\pi}C_F
\left(\frac{4\pi\mu^2}{Q^2}\right)^\epsilon\frac{1}{\Gamma(1-\epsilon)}
\int\frac{dx}{x}\frac{dx'}{x'} T_F(x,x)\bar q(x')C_F\delta(1-z)
\left[-\frac{2}{\epsilon^2}-\frac{3}{\epsilon}-8+\pi^2\right]\, .
\label{virt}
\end{eqnarray}
In the next section, we will calculate the real-gluon radiation contribution,
and obtain the final result for the NLO correction. It is important to verify
that the soft divergence in the above virtual corrections is canceled
against that in the real diagrams. We will check this cancelation in
the next section.

\section{Real Corrections and Final Results}

The real-gluon emission contributions to the single-spin asymmetry for
the Drell-Yan process have been computed in~\cite{jqvy}. In that paper,
the focus was on the SSA at fixed $q_\perp$, whereas in our present
calculation we are considering the $q_\perp$-weighted cross section,
which involves integration over all $q_\perp$. While we can still
use much of the set-up of the calculations of~\cite{jqvy}, we have to
redo them in $D=4-2\epsilon$ dimensions. This is relatively
straightforward. Another issue we need to address is the
transverse-momentum flow in the diagrams. In the calculations
performed in~\cite{jqvy}, the virtual photon is an ``observed'' particle
in the sense that its momentum is kept fixed in the collinear expansion of
the hard partonic scattering amplitudes $H(k_{q1},k_{q2};Q^2)$. Since
we are integrating over all $q_\perp$ in the present calculation, one
might think that the photon momentum could be a function of the
transverse momenta of the initial partons and hence change the collinear
expansion. However, because of momentum conservation it turns out
that it is sufficient to allow transverse-momentum flow only through
the radiated gluon.

Before going into the calculation, we note that we will only consider
contributions associated with the twist-three quark-gluon correlation
function. There are two types of such contributions:
real-gluon emission in the subprocess $(qg)+\bar{q}\to \gamma^*+g$,
and the quark-gluon Compton process $(qg)+g\to \gamma^*+q$. We do not
consider in this study contributions entering with a three-gluon
twist-three correlation function~\cite{new1,Ji:1992eu}. We also ignore contributions
by the ``axial'' twist-three quark-gluon correlation function
considered in Ref.~\cite{KVY}.

The real-gluon radiation diagrams yield soft and collinear divergences
when integrated over the transverse momentum, which are regularized
by dimensional regularization. Following a standard procedure for the phase
space integrals as in the case of the spin-averaged cross section~\cite{dy},
we obtain the following expression for the real-gluon radiation contribution
to the transverse momentum weighted spin-dependent cross section:
\begin{eqnarray}
\frac{d \langle q_\perp\Delta \sigma (S_\perp) \rangle}{dQ^2}
&=&\sigma_0\frac{\alpha_s}{2\pi}\int\frac{dx}{x}\frac{dx'}{x'}
\left(\frac{4\pi\mu^2}{Q^2}\right)^\epsilon\frac{1}{\Gamma(1-\epsilon)}
z^\epsilon(1-z)^{1-2\epsilon}\int_0^1 dv (v(1-v))^{-\epsilon}\nonumber\\
&&
\left\{x\frac{\partial}{\partial x}T_F(x,x)\left(D_{q\bar q} \bar q(x')
+D_{qg} g(x')\right)\nonumber\right.\\
&&+T_F(x,x)\left(N_{q\bar q}^s \bar q(x')+N_{qg}^s g(x')\right)\nonumber\\
&&\left.+T_F(x,x-\bar x_g)\left(N_{q\bar q}^h \bar q(x')+N_{qg}^h g(x')
\right)\right\} \ ,
\end{eqnarray}
where $v$ is related to the partonic center-of-mass scattering angle $\theta$
by $v=(1+\cos\theta)/2$. The above expression contains three contributions.
The first two are the derivative and non-derivative terms from
the soft-pole diagrams, respectively; the third is the contribution
by hard-pole diagrams, which only have non-derivative pieces.
In $D=4-2\epsilon$ dimensions, we obtain the following expressions
for the corresponding partonic hard-scattering terms~\footnote{Note that
in contrast to~\cite{dy}
we average over the polarizations of the initial gluon in the $qg$
subprocess by the factor $1/(2(1-\epsilon))$, as is customary
in the $\overline{{\mathrm{MS}}}$ scheme.}:
\begin{eqnarray}
D_{q\bar q}&=&\frac{1}{2N_c}\frac{-\hat t}{\hat s}
\left[(1-\epsilon)\left(\frac{\hat u}{\hat t}+
\frac{\hat t}{\hat u}\right)+\frac{2Q^2\hat s}{\hat t\hat u}-
2\epsilon\right] \ , \\
D_{qg}&=&-\frac{N_c^2}{2(N_c^2-1)}\frac{-\hat t}{\hat s}
\left[(1-\epsilon)\left(\frac{\hat s}{-\hat t}+
\frac{-\hat t}{\hat s}\right)-\frac{2Q^2\hat u}{\hat t\hat s}+
2\epsilon\right] \ ,\\
N_{q\bar q}^{(s)}&=&\frac{1}{2N_c}\frac{1}{-\hat s\hat t\hat u}
\left[Q^2(\hat u^2-\hat t^2)+2Q^2\hat s
(Q^2-2\hat t)-(\hat u^2+\hat t^2)\hat t\nonumber\right.\\
&&\left.+\epsilon(\hat s\hat t^2-\hat s\hat u^2+2\hat t^3+3
\hat t^2\hat u-\hat u^3)\right]\ , \\
N_{qg}^{(s)}&=&\frac{N_c^2}{2(N_c^2-1)}\frac{1}{-\hat s^2\hat t}
\left[Q^2(\hat s^2-\hat t^2)+2Q^2\hat u
(Q^2-2\hat t)-(\hat s^2+\hat t^2)\hat t\nonumber\right.\\
&&\left.+\epsilon(\hat u\hat t^2-\hat u\hat s^2+2\hat t^3+3\hat t^2
\hat s-\hat s^3)\right]\ , \\
N_{q\bar q}^{(h)}&=&\left(\frac{1}{2N_c}+C_F\frac{\hat s}
{\hat s+\hat u}\right)
\frac{(Q^2-\hat t)^3+Q^2\hat s^2-\epsilon(\hat s+\hat u)
(\hat t+\hat u)\hat u}{\hat s\hat t\hat u} \ , \\
N_{qg}^{(h)}&=&\left(\frac{-N_c^2}{2(N_c^2-1)}+T_R\frac{\hat s}
{\hat s+\hat u}\right)
\frac{(Q^2-\hat t)^3+Q^2\hat u^2-\epsilon(\hat s+\hat u)
(\hat s+\hat t)\hat s}{-\hat s^2\hat t} \ ,
\end{eqnarray}
where $\hat s$, $\hat t$, and $\hat u$ are the partonic Mandelstam
variables for the $2\to 2$ processes, which can be expressed in terms
of $Q^2$ and $z$ as
\begin{equation}
\hat s=\frac{Q^2}{z},~~\hat t=-\frac{Q^2}{z}(1-z)(1-v),~~
\hat u=-\frac{Q^2}{z}(1-z)v \ .
\end{equation}
 From the above expressions, we see that the integral over $v$ will contain
divergences when $v\to 0$ or $v\to 1$. The main task in this calculation
is to separate these divergences and identify them as soft or collinear,
so that they can be canceled appropriately.

First, let us examine the derivative term from the soft-gluon pole
contribution in the $q\bar q$ channel. After integrating over $v$,
it becomes
\begin{eqnarray}
&&\sigma_0\frac{\alpha_s}{2\pi}\left(\frac{4\pi\mu^2}
{Q^2}\right)^\epsilon\frac{1}{\Gamma(1-\epsilon)}\frac{1}{2N_c}
\int\frac{dx}{x}\frac{dx'}{x'}\bar q(x')\left\{
\left(-\frac{1}{\epsilon}\right)T_F(x,x)\, 2 z^2+\right.\nonumber\\[3mm]
&&\left. ~~~~~~~~~~~\times
\left[x\frac{\partial}{\partial x}T_F(x,x)\right] (1+z^2)
\ln\frac{(1-z)^2}{z}\right\} \ ,
\end{eqnarray}
where we have performed an integration by parts in order to simplify
the $1/\epsilon$ term. The latter comes from a collinear divergence, which
will be canceled by factorization into the evolved $T_F$ function.
There is no soft divergence in this term, which is expected because
the virtual diagrams do not contribute to the derivative terms, as we saw
earlier.

The soft-pole derivative terms in the $qg$ channel give only finite
contributions. It is easy to perform the phase space integration, and
we obtain
\begin{equation}
\sigma_0\frac{\alpha_s}{2\pi}\frac{-N_c^2}{2(N_c^2-1)}
\int\frac{dx}{x}\frac{dx'}{x'}
\left[x\frac{\partial}{\partial x}T_F(x,x)\right] g(x') (1+z^2)
\left[\frac{1}{3}(1-z)(4+4z^2-5z)\right] \ .
\end{equation}

The non-derivative term from the soft-pole diagrams in the
$q\bar q$ channel has both soft and collinear divergences.
After integrating over the phase space variable $v$, we get
\begin{eqnarray}
&&\sigma_0\frac{\alpha_s}{2\pi}\left(\frac{4\pi\mu^2}{Q^2}\right)^\epsilon
\frac{1}{\Gamma(1-\epsilon)}\frac{1}{2N_c}
\int\frac{dx}{x}\frac{dx'}{x'}T_F(x,x) \bar q(x')
\left\{-\frac{2}{\epsilon^2}\delta(1-z)
-\frac{1}{\epsilon}\frac{z^3-3z^2-z-1}{(1-z)_+}
\nonumber\right.\\[2mm]
&&\left.+\frac{\pi^2}{3}\delta(1-z)+(z^3-3z^2-z-1)
\left[2\left(\frac{\ln(1-z)}{1-z}\right)_++\frac{\ln z}{1-z}\right]\right\}\ .
\label{real1}
\end{eqnarray}
The double-pole $1/\epsilon^2$ term represents a soft-collinear
divergence, which will eventually be canceled. The collinear-divergent
term $\propto 1/\epsilon$ will generate part of the splitting function for
the evolution of the unpolarized quark distribution and/or the
twist-three correlation function.

Now we turn to the hard-pole contributions, which are only non-derivative.
For these contributions, the two arguments in the twist-three quark-gluon
correlation function are different and may depend on partonic kinematics.
As a result, the $v$-integral is somewhat more involved, and for some
parts the integral cannot be performed completely. First, we will
separate these parts by introducing ``plus''-distributions of the form
\begin{equation}
\int_0^1 dv \frac{g(v)}{v_+}\equiv\int dv \frac{g(v)-g(0)}{v}\ ,~~~
\int_0^1 dv \frac{g(v)}{(1-v)_+}\equiv\int dv \frac{g(v)-g(1)}{1-v}\ ,
\end{equation}
in the integrand. The distributions arise
from terms $\propto v^{-1}, (1-v)^{-1}$ in the integrand which, when
combined with the phase space factor $(v(1-v))^{-\epsilon}$, give rise
to identities of the form
\begin{equation}
v^{-1-\epsilon}=-\frac{1}{\epsilon} \delta(v) + \frac{1}{(v)_+} -
\epsilon \left(\frac{\ln(v)}{v}\right)_+ + {\cal O}(\epsilon^2) \; .
\end{equation}
For the $q\bar q$ channel, the part that cannot be further integrated
over $v$ analytically then reads:
\begin{eqnarray}
&&\sigma_0\frac{\alpha_s}{2\pi}\int\frac{dx}{x}\frac{dx'}{x'}\frac{dv}{1-z}
\left(\frac{1}{v_+}+\frac{1}{(1-v)_+}\right)T_F\left(x,x\frac{z}{1-v(1-z)}
\right)
\bar q(x')
\nonumber\\
&&~~~~~\times \left[(1-v(1-z))^3+z\right]
\left(\frac{1}{2N_c}+C_F\frac{1}{1-v(1-z)}\right) \ .
\end{eqnarray}
Both plus distributions are needed because the integrand is divergent at
both $v\to 0$ and $v\to 1$. The distributions guarantee that the integral
over $v$ is finite. Besides, as one can see, there is no divergence
in the limit $z\to 1$. The remaining part of the hard-pole contribution
in the $q\bar q$ channel is then rather straightforward to obtain,
and we obtain upon integration over $v$
\begin{eqnarray}
&&\sigma_0\frac{\alpha_s}{2\pi}
\left(\frac{4\pi\mu^2}{Q^2}\right)^\epsilon\frac{1}{\Gamma(1-\epsilon)}
\int\frac{dx}{x}\frac{dx'}{x'} \bar q(x')
\left\{T_F(x,xz)\left(\frac{1}{2N_c}+C_F\right)
\left[\frac{1}{\epsilon^2}\delta(1-z)
\nonumber\right.\right.\\
&&~~~~~\left.-\frac{1}{\epsilon}\frac{1+z}{(1-z)_+}
-\frac{\pi^2}{6}\delta(1-z)+(1+z)
\left(2\left(\frac{\ln(1-z)}{1-z}\right)_++\frac{\ln z}{1-z}
\right)\right]\nonumber\\
&&+T_F(x,x)\left(\frac{1}{2N_c}z+C_F\right)\left[
\frac{1}{\epsilon^2}\delta(1-z)-\frac{1}{\epsilon}\frac{1+z^2}{(1-z)_+}
    -\frac{\pi^2}{6}\delta(1-z)\nonumber\right.\\
&&\left.\left.~~~~~+(1+z^2)\left(2\left(\frac{\ln(1-z)}{1-z}\right)_++
\frac{\ln z}{1-z}\right)+(1-z)\right]\right\}\ .
\label{real2}
\end{eqnarray}
Again, we have both soft and collinear divergences.

Similarly, for the hard-pole contribution in the $qg$ channel,
we have a regularized part that cannot be further integrated
over $v$ analytically:
\begin{eqnarray}
&&\sigma_0\frac{\alpha_s}{2\pi}\int\frac{dx}{x}
\frac{dx'}{x'}\frac{dv}{(1-v)_+}T_F\left(x,x\frac{z}{1-v(1-z)}\right)
g(x') \nonumber\\
&&~~~~~\times \left[(1-v(1-z))^3+v^2(1-z)^2z\right]
\left(\frac{-N_c^2}{2(N_c^2-1)}+T_R\frac{1}{1-v(1-z)}\right) \ ,
\end{eqnarray}
and a remaining singular part:
\begin{eqnarray}
&&\sigma_0\frac{\alpha_s}{2\pi}\left(\frac{4\pi\mu^2}{Q^2}\right)^\epsilon
\frac{1}{(1-\epsilon)\Gamma(1-\epsilon)}
\int\frac{dx}{x}\frac{dx'}{x'} T_F(x,x)g(x')
\left(\frac{-N_c^2}{2(N_c^2-1)}z+T_R\right)\nonumber\\[2mm]
&&~~~~~\times\left[-\frac{1}{\epsilon}
\left(z^2+(1-z)^2\right)+(z^2+(1-z)^2)\ln\frac{(1-z)^2}{z}+1\right]\ .
\end{eqnarray}
There is no soft divergence. The collinear divergence will
be canceled by factorization of the gluon splitting contribution to the
spin-averaged anti-quark distribution function.

As we mentioned at the beginning, the soft divergence has to disappear
after adding the contributions by the real-gluon radiation and virtual
diagrams. Indeed this happens, as inspection of Eqs.~(\ref{virt}) and
(\ref{real1}),~(\ref{real2}) shows. Specifically, the
term $\propto 1/\epsilon^2$ from the soft-pole diagrams cancels
that from the hard-pole diagrams associated with color-factor $1/2N_c$, and
the remaining $1/\epsilon^2$ term from the hard-pole diagrams (associated
with the color-factor $C_F$) cancels against that from the virtual
diagrams. This is an important cross-check on the consistency of our
calculations, and demonstrates the importance of the hard-pole diagrams.

After cancelation of soft poles, the result will only contain
collinear divergences. We find for the remaining pole term
\begin{eqnarray}
&&\sigma_0\frac{\alpha_s}{2\pi}
\int\frac{dx}{x}\frac{dx'}{x'} \left(-\frac{1}{\epsilon}\right)\left\{
T_F(x,x)\bar q(x')\left[2C_F\left(\frac{1+z^2}{1-z}\right)_++
(C_F+\frac{1}{2N_c})z\right]\right.\nonumber\\
&&+(T_F(x,xz)-T_F(x,x))\bar q(x')(C_F+\frac{1}{2N_c})\frac{1+z}{1-z}\nonumber\\
&&\left.+T_F(x,x)g(x')T_R(z^2+(1-z)^2)\right\}\ .
\end{eqnarray}
The residue of this collinear divergence contains the splitting functions
governing the evolution of the anti-quark distribution in the unpolarized
nucleon and the twist-three correlation function. For the former, we have
\begin{eqnarray}
\bar q(x)=\bar q^{(0)}(x)+\frac{\alpha_s}{2\pi}\int \frac{dx'}{x'}
\left(-\frac{1}{\epsilon}\right)
\left[\bar q(x')C_F\left(\frac{1+z^2}{1-z}\right)_++g(x')
T_R(z^2+(1-z)^2)\right] \ ,
\end{eqnarray}
where $z=x/x'$ and $\bar q^{(0)}(x)$ denotes the ``bare'' leading order
anti-quark distribution. Similarly, we obtain the collinear QCD
correction to the Qiu-Sterman matrix element at equal momentum fractions:
\begin{eqnarray}
T_F(x,x)&=&T_F^{(0)}(x,x)+\frac{\alpha_s}{2\pi}\int \frac{dx'}{x'}
\left(-\frac{1}{\epsilon}\right)
\left\{T_F(x',x')C_F\left(\frac{1+z^2}{1-z}\right)_+\nonumber\right.\\
&&+\left. \left(C_F+\frac{1}{2N_C}\right)\left[\frac{1+z}{1-z}T_F(x',x'z)-
\frac{1+z^2}{1-z}T_F(x',x')\right] \right\} \ .
\end{eqnarray}
 From this equation, we directly read off the scale evolution
equation for the ``diagonal'' twist-three quark-gluon correlation
function at $x_1=x_2=x$:
\begin{eqnarray} \label{evonew}
\frac{\partial}{\partial \ln \mu^2}T_F(x,x;\mu^2)&=&
\frac{\alpha_s(\mu^2)}{2\pi}
\int \frac{dx'}{x'}\left\{C_F T_F(x',x';\mu^2)
\left(\frac{1+z^2}{1-z}\right)_+\right.\\
&&\left.+\frac{N_c}{2} \left[T_F(x',x';\mu^2)z-(T_F(x',x';\mu^2)
-T_F(x',x'z;\mu^2))\frac{1+z}{1-z}\right]\right\} \nonumber\\
&&\equiv \frac{\alpha_s(\mu^2)}{2\pi}\int\frac{dx'}{x'} {\cal P}_{qg\to qg}
\otimes T_F(x',x';\mu^2)\ .
\end{eqnarray}
This evolution equation could also have been obtained from the 
perturbative calculation of the quark Sivers function at large 
transverse momentum performed in~\cite{jqvy}. However,  
we note that in that paper a boundary term ($\propto \delta(1-z)$) 
in the derivative contribution was overlooked\footnote{This term
does not, however, affect the consistency of the twist-three and
the TMD approaches established in~\cite{jqvy}.}.
After correcting for this term, the result of~\cite{jqvy} becomes 
consistent with that given above. Eq.~(\ref{evonew}) is also consistent 
with the results derived recently
by different methods~\cite{kang,zhou}. We note, however, that the evolution
equations derived in~\cite{kang} go beyond ours, as they also contain
the contributions from additional operators such as three-gluon ones,
which we are not considering here.

A few comments on the evolution equation
are in order. First, it is evident that the scale evolution of the
``diagonal'' ($x_1=x_2=x$) function mixes with the function at
$x_1\neq x_2$, implying that the equation as it stands is not
closed. In other words, there will be a more general evolution
equation for the full function $T_F(x_1,x_2)$. This feature is
quite general for higher-twist parton distributions and fragmentation
functions~\cite{Ratcliffe:1985mp,Ali:1991em,{Belitsky:1997zw},belitsky}. Second, there is no particular
simplification of the evolution equation in the large-$N_c$ limit. This is
different from what was discovered for the evolution equations for
other twist-three quark distributions, such as  $h_L(x)$ and $e(x)$,
where the evolution equations are closed (diagonal) in that
limit~\cite{Ali:1991em}. However, we notice that the high-$x$
part of the evolution equation, i.e., the large $z$ limit of the kernel
in the integrand, is the same as that for the spin-averaged leading-twist
quark distribution, because the term $C_F ((1+z^2)/(1-z))_+$ is the
ordinary leading order quark splitting function. This property will
have important phenomenological consequences~\cite{kang} for the behavior of
the quark-gluon correlation function at high $x$ and thus for
SSAs in hadronic processes at forward angles.

We note that the renormalization and evolution of general twist-three
quark-gluon operators has been extensively studied over the past
two decades~\cite{Ratcliffe:1985mp,Ali:1991em,{Belitsky:1997zw},belitsky}.
The above result for the evolution
of the Qiu-Sterman matrix element should likely also be reproduced
from the evolution equations discussed in some of these papers.
However, we notice that the Qiu-Sterman matrix element corresponds
to a very different projection of the general twist-three quark-gluon
correlation function, and we do not expect that there will be a simple
relation between the above evolution equation and those for other
specific twist-three distributions such as $g_T(x)$. The comparison
between the above results (or the ones of~\cite{kang,zhou}) and those
in~\cite{Ratcliffe:1985mp,Ali:1991em,{Belitsky:1997zw},belitsky} is
very important and will be addressed in the future.

After $\overline{\rm MS}$ subtraction of the collinear divergences
into the quark-gluon correlation function of the polarized nucleon
and the anti-quark distribution of the unpolarized nucleon,
we obtain the full NLO expression for the soft-gluon and hard-pole 
contributions\footnote{We remind the reader that we do not consider 
contributions associated with soft-fermion poles or with a 
three-gluon twist-three correlation function.} to the 
transverse-momentum weighted single-spin dependent cross section in 
Drell-Yan lepton pair production in $pp$ collisions:
\begin{eqnarray}
\frac{d \langle q_\perp\Delta \sigma (S_\perp) \rangle}{dQ^2}
&=&\sigma_0\int\frac{dx}{x}\frac{dx'}{x'} T_F(x,x;\mu^2)
\bar q(x';\mu^2)\nonumber\\
&&\hspace*{-2cm}
+\sigma_0\frac{\alpha_s}{2\pi}\int\frac{dx}{x}\frac{dx'}{x'}
\left\{\bar q(x';\mu^2)
\left[\ln\frac{Q^2}{\mu^2}\left(C_F{\cal P}_{qq}+
{\cal P}_{qg\to qg}\otimes T_F(x,xz;\mu^2)
\right)\nonumber\right.\right.\\
&&\hspace*{-2cm}+\frac{1}{2N_c}(x\frac{\partial}{\partial x}
T_F(x,x;\mu^2))(1+z^2)\ln\frac{(1-z)^2}{z}
+\left(2\left(\frac{\ln(1-z)}{1-z}\right)_+
-\frac{\ln z}{1-z}\right)\nonumber\\
&&\hspace*{-2cm}\times \left((C_F(1+z^2)+
\frac{2z^3-3z^2-1}{2N_c})T_F(x,x;\mu^2)+
(\frac{1}{2N_c}+C_F)(1+z)T_F(x,xz;\mu^2)\right)\nonumber\\
&&\hspace*{-2cm}\left.+T_F(x,x;\mu^2)\left((C_F+\frac{z}{2N_c})(1-z)+
C_F(\frac{2\pi^2}{3}-8)\delta(1-z)\right)\right]\nonumber\\
&&\hspace*{-2cm}+g(x';\mu^2)\left[(x\frac{\partial}{\partial x}T_F(x,x;\mu^2))
(\frac{-N_c^2}{2(N_c^2-1)})\frac{1}{3}(1-z)(4+4z^2-5z)\right.\nonumber\\
&&\hspace*{-2cm}+T_F(x,x;\mu^2)T_R\left((z^2+(1-z)^2)\ln\frac{Q^2}{\mu^2}
\frac{(1-z)^2}{z}+2z(1-z)\right)\nonumber\\
&&\hspace*{-2cm}\left.\left.+T_F(x,x;\mu^2)\frac{N_c^2}{2(N_c^2-1)}
\frac{1}{6}(8-27z+48z^2-29z^3)\right]\right\}\nonumber\\
&&\hspace*{-2cm}+\int\frac{dv}{1-z} \left(\frac{1}{v_+}+
\frac{1}{(1-v)_+}\right)T_F(x,x\frac{z}{1-v(1-z)};\mu^2) \bar q(x';\mu^2)
\nonumber\\
&&\hspace*{-2cm}\times \left[(1-v(1-z))^3+z\right]
\left(\frac{1}{2N_c}+C_F\frac{1}{1-v(1-z)}\right) \nonumber\\
&&\hspace*{-2cm}+\int\frac{dv}{(1-v)_+}T_F(x,x\frac{z}{1-v(1-z)};\mu^2)
g(x';\mu^2)
\left[(1-v(1-z))^3+v^2(1-z)^2z\right]\nonumber\\
&&\hspace*{-2cm}\times \left(\frac{-N_c^2}{2(N_c^2-1)}+
T_R\frac{1}{1-v(1-z)}\right) \ .
\end{eqnarray}
As expected, the logarithms containing the factorization scale
enter with the splitting functions for the evolution of the
twist-three quark-gluon correlation function and the twist-two
anti-quark distribution.

One important feature of this result is its behavior near ``partonic
threshold'', that is in the large-$z$ limit of the integrand,
corresponding to $\hat{s}\sim Q^2$, when the initial partons
have ``just enough'' energy to produce the virtual photon.
Setting the scale $\mu=Q$, we have the following structure of
the NLO correction in this case:
\begin{eqnarray}
\frac{d \langle q_\perp\Delta \sigma (S_\perp) \rangle}{dQ^2}
&=&\sigma_0\frac{\alpha_s}{2\pi}\int\frac{dx}{x}\frac{dx'}{x'}
T_F(x,x;\mu^2) \bar q(x';\mu^2)\left[
8C_F\left(\frac{\ln(1-z)}{1-z}\right)_+ +\ldots \right].\nonumber\\
\end{eqnarray}
Here we have written out the ``double-logarithmic'' term which
dominates near threshold in the $\overline{\rm MS}$ scheme.
The ellipses denote terms that are
subleading relative to this term. The structure of this expression
is identical to that for the spin-averaged $q_\perp$-integrated
NLO cross section near threshold,
\begin{eqnarray}
\frac{d \sigma}{dQ^2}
&=&\sigma_0\frac{\alpha_s}{2\pi}\int\frac{dx}{x}\frac{dx'}{x'} q(x;\mu^2)
\bar q(x';\mu^2)\left[
8C_F\left(\frac{\ln(1-z)}{1-z}\right)_+ +\ldots \right] \ .
\end{eqnarray}
This means that the soft gluon contribution is spin-independent.
It contributes in the same way to the spin-averaged and single-spin-dependent
cross sections, and will lead to the same soft-gluon threshold resummation
effects to these cross sections, at least at the leading double
logarithmic level. This observation is very similar to that made
for the transverse momentum resummation in the Drell-Yan
process~\cite{Idilbi:2004vb}.


\section{Conclusions}

In summary, we have derived the NLO perturbative-QCD correction
to the transverse momentum weighted single spin asymmetry
in Drell-Yan lepton pair production in hadronic collisions.
In the calculation, we have shown that the collinear divergences
can be absorbed into the NLO twist-three quark-gluon correlation
function of the transversely polarized nucleon and the unpolarized
quark distribution of the unpolarized nucleon. This procedure
also determines the evolution equation for the ``diagonal part''
of the twist-three Qiu-Sterman matrix element at equal momentum
fractions, $x_1=x_2$. We have found this equation to be consistent
with the more complete one derived recently in Ref.~\cite{kang,zhou}.

Our calculations suggest that a general factorization formula
(see Eq.~(3)) exists for the transverse momentum weighted
spin-dependent cross section in the Drell-Yan process, in
extension of the general factorization arguments given
in~\cite{qiu-fac}.

We have found that both the evolution kernel and the full NLO expression
for the spin-dependent cross section become identical to their
spin-averaged counterparts in the ``threshold'' limit
$\hat{s}\to Q^2$, or $z\to 1$. This will likely have the phenomenological
consequence that the single-spin asymmetry for the Drell-Yan process
will be quite stable under NLO corrections, in particular when
$\tau=Q^2/s$ is large.

It will be important to carry out further studies. We have mentioned
already that it may be possible to derive the evolution of the twist-three
correlation functions also from some of the results
of~\cite{Ratcliffe:1985mp}. Also, it will be important to derive
NLO corrections also for other processes. For example, extension to
semi-inclusive deep inelastic scattering should be
relatively straightforward to do.

\section*{Acknowledgments}
We thank Zhongbo Kang, Jianwei Qiu, and Jian Zhou for useful comments and
valuable discussions. W.V. is grateful to V. Braun, M. Diehl, and
D. M\"{u}ller for useful discussions.
This work was supported in part by the U.S.
Department of Energy under grant contract DE-AC02-05CH11231. F.Y.
and W.V. thank RIKEN, Brookhaven National Laboratory and the U.S.
Department of Energy (contract number DE-AC02-98CH10886) for
providing the facilities essential for the completion of their
work.

\end{document}